\documentclass[12pt]{article}
\usepackage[cp1251]{inputenc}
\usepackage[T2A]{fontenc}
\usepackage[dvips]{graphicx}
\usepackage{latexsym}
\usepackage{amssymb}

\begin{document}

\title{On properties of the distribution of virtual wormholes in a vacuum}
\author{A. A. Kirillov, E. P. Savelova \\
Bauman Moscow State Technical University, \\
Moscow, 105005, Russian Federation
}
\date{}
\maketitle

\begin{abstract}
A model of space-time foam in the form of an arbitrary distribution of spherical Euclidean wormholes is considered. A method for constructing the exact solution of Einstein's Euclidean equations for the metric corresponding to this model is proposed. In the framework of our model we obtain the expression for the Euclidean action and its dependence on the parameters of wormholes in the explicit form. It is shown how the solutions obtained make it possible to determine all possible correlation functions associated with the parameters of virtual wormholes in the vacuum state.

\end{abstract}

\section{Introduction}
Modeling the effects of space-time foam is of considerable interest in particle physics. In particular, it is expected that these effects should lead to natural dynamic cutoff at ultra-small (sub-Planck) scales and thereby eliminate ultraviolet divergences in quantum field theory. Indeed, the first indications of such behavior were obtained in two-dimensional models of quantum gravity \cite{frac,frac1} where it was found that on Planck scales, the space-time foam has fractal properties. Lattice models of quantum gravity also indicate the fractality of the topological structure of spacetime on small scales. In particular, the spectral dimension of space-time was shown  to depend on the scale \cite{frac2}.
%%%%%%%%%%

However, the lattice models are based primarily on numerical research and correspond to a deeply nonperturbative domain. In this area of scales, the concept of space-time itself is absent in the usual sense. Accordingly, it is quite difficult to establish a correspondence with the description of the effects of space-time foam within the framework of perturbative methods, i.e., in the area of scales where the background space is already well distinguished and it is possible to talk about different particles. In this region of scales, the model in which the space-time foam is modeled by the presence of an arbitrary number of virtual wormholes seems to be the most adequate.

The virtual wormhole corresponds to the fluctuation of the topology of space describing the virtual process
when  some daughter universe separates from our mother Universe and afterwards re-joins back onto the mother Universe. Virtual processes correspond to the classically forbidden areas of the configuration space (located under a potential barrier). Accordingly, such metrics are described by Euclidean configurations.

%%%%%%%%%%%%%%%%%%%
It turns out that the scattering of ordinary particles on virtual wormholes leads to the generation of an infinite series of additional excitations corresponding to very heavy particles \cite{KSPV}. In other words, the presence of virtual wormholes automatically leads to the dynamic implementation of the invariant Pauli-Villars regularization scheme \cite{PV1,PV2} with an infinite number of auxiliary fields. However, in contrast to the Pauli-Villars scheme, auxiliary particles have the status of real particles here. Moreover, since auxiliary particles and fields have an origin associated with the non-triviality of the topology of space-time, such fields are generated for all types of fundamental particles.

%%%%%%%%%%%%%%%%%
It is also expected that the space-time foam should predict a number of new phenomena and effects in particle physics. The most interesting are the effects associated with the formation of real (not virtual) wormholes and similar objects. In other words, processes in which the real topological structure of space can change are of particular interest. Indeed, additional auxiliary fields generated by virtual wormholes predict the presence of a series of phase transitions in the early universe. According to the standard Kibble \cite{Kibble} scenario, phase transitions are accompanied by the formation of macroscopic defects such as domain walls. Some of these walls carry negative energy and, due to their macroscopic nature, such walls can serve as a source for the formation and stability support of real wormholes \cite{KS23}. The possibility of recreating such conditions in the laboratory is also of undoubted interest \cite{KS22}.

%%%%%%%%%%%%%%%%%%%
Rigorous calculation of various field correlation functions requires knowledge of the vacuum distribution of virtual wormholes and various average values of the wormhole parameters. At first glance, this also requires numerical investigation. However, in this paper we show that the properties of the vacuum distribution of virtual wormholes allow analytical investigation, at least in the low-energy limit.

\section{Vacuum $n$-point wormhole distribution functions}

In quantum field theory, in the path integral approach, all correlation functions can be obtained from the generating function $Z(J)$, which is defined as follows
\begin{equation}
Z_{tot}\left( J\right) =\sum_{\tau }\int e^{-S_{E}}D\varphi ,
\end{equation}%
where the sum is taken over topologies $\tau$, and the integral is taken over configurations of fields $\varphi$.
 %%%%%%%
Here $S_{E}$ is a Euclidean action in which it is convenient to separate the background geometry $h$, while deviations from the background metric or perturbations are assumed to be small and will be considered along with all other fields of matter $\varphi$. The background geometry depends on the topological structure of the space $h(\tau)$. We will assume that the topology is given by the distribution of an arbitrary number of wormholes. To explicitly calculate the sum over topologies, we introduce a distribution function, that is, the density of the distribution of wormholes in the configuration space.
\begin{equation}
F\left( \xi \right) =\sum_{A=1}^{N}\delta \left( \xi -\xi _{A}\right)
\label{F}
\end{equation}%
where $\xi _{A}=\left(a_{A},R_{A}^{+},R_{A}^{-},\Lambda _{\beta A}^{\alpha}\right) $ is a set of parameters that define the Euclidean wormhole and its position in space. Then the sum over the topologies is
\begin{equation}
\sum_{\tau }=\sum_{N=0}^{\infty }\int DF
\end{equation}%
where $DF=\prod_{A=1}^{N}d\xi _{A}$. The action is taken in the form of
\begin{equation}
S_{E}=S_{E}\left( h\right) +\frac{1}{2}\left( \varphi ,\left( -\Delta
+m^{2}\right) \varphi \right) -\left( J,\varphi \right)
\end{equation}%
where $S_{E}\left(h\right)$ is determined by the background geometry (in the explicit form, the action for the background geometry is given by the expression (\ref{Act})), $\varphi$ corresponds to all fields defined on the background space $E^{\ast}$ (which also include and gravitational perturbations).  We also introduced the notation $\left(J,\varphi\right) =\int_{E^{\ast}}J\left(x\right) \varphi\left(x\right) \sqrt{g}d^{4}x$.
Integration by field configurations gives
\begin{equation}
Z^{\ast }\left( J,\xi \right) =Z_{0}\left( G\right) e^{\frac{1}{2}\left(
J,GJ\right) -S_{E}\left( h\right) },
\end{equation}%
where $Z_{0}(G) =\int e^{-\frac{1}{2}\left( \varphi ,\left( -\Delta +m^{2}\right) \varphi\right) }D\varphi $, and $G=G\left(\xi_{1},...\xi_{N}\right) $ is the Green function for a fixed background topology (i.e., a fixed distribution of wormholes).
In the future, it is convenient to use the following decomposition for Green's function (see, for example, \cite{S15})
$$
G\left(\xi _{1},...\xi _{N}\right) =G_{0}+\delta G\left( \xi _{1},...\xi _{N}\right) ,
$$
where $G_0$ corresponds to the standard Green function in Euclidean space, and the term $\delta G=\sum\delta G_{n}$ describes corrections related to scattering on wormholes. So $\delta G_{n}$ defines n-fold scattering on wormholes.
A single scattering is described by a term of the form
$$
\delta G_{1}=\int \delta G_{1}\left( x,x^{\prime },\xi \right) F\left( \xi \right) d\xi ,
 $$
where $F$ is the distribution function defined in (\ref{F}). Accordingly, multiple scattering corresponds to terms  of the form
$$
\delta G_{n}=\int \delta G_{n}\left( \xi_1,...,\xi_n \right) \prod_{k=1}^n F( \xi_k )d\xi_k .
 $$

The original generating function $Z_{tot}(J)$ is defined by the integral
\begin{equation}
Z_{tot}(J) =\int Z_{0}(G) e^{\frac{1}{2}\left(
J,GJ\right) -S_{E}\left( h\right) }DF.
\end{equation}%
Then, expanding out $Z_{tot}(J)$ in a series by small $J$, we get
 \cite{KSPV}
\begin{equation}
W\left( J\right) =\frac{1}{2}\overline{\left( J,GJ\right) }+\frac{1}{8}%
\overline{\left( J,\Delta GJ\right) ^{2}}+\frac{1}{48}\overline{\left(
J,\Delta GJ\right) ^{3}}+...  \label{W}
\end{equation}%
where $W\left( J\right) =\ln \frac{Z_{tot}\left( J\right) }{Z_{tot}\left(
0\right) }$, $\Delta G=G-\overline{G}$,
and the overline denotes the corresponding average values for the vacuum, for example,
\begin{equation}\label{Mn}
\overline{G}=\left\langle 0\left\vert G\right\vert 0\right\rangle _{J=0}=%
\frac{1}{Z_{tot}\left( 0\right) }\int GZ_{0}\left( G\right) e^{-S_{E}\left(
h\right) }DF.
\end{equation}%
All vacuum averages in (\ref{W}) can be expressed in terms of vacuum averages from n-point distributions of wormholes in the configuration space.  Indeed, we define n-point distributions as
\begin{equation}\label{rho}
\rho _{n}\left( \xi _{1},...\xi _{n}\right) =\left\langle 0\left\vert
\prod\limits_{i=1}^{n}F\left( \xi _{i}\right) \right\vert 0\right\rangle
_{J=0},
\end{equation}%
where $F$ is given by (\ref{F}). Then we find
\begin{equation}
\overline{G}=G_{0}+\sum_{n=1}^{\infty }\int \delta G_{n}\left(
\xi _{1},...\xi _{n}\right) \rho _{n}\left( \xi _{1},...\xi _{n}\right) d^{n}\xi
.
\end{equation}%
Similarly, we get
\begin{equation}
\overline{\Delta G\Delta G}=\sum_{m,n=1}^{\infty }\int \delta G_{n}\left(
\xi \right) \delta G_{m}\left( \xi ^{\prime }\right) \Delta \rho
_{n,m}\left( \xi ,\xi ^{\prime }\right) d^{n}\xi d^{m}\xi ^{\prime }
\end{equation}%
where we denote
\begin{equation}\label{DR}
\Delta \rho _{n,m}\left( \xi _{1},...,\xi _{n},
\xi _{1}^{\prime },...,\xi _{m}^{\prime }\right) =\rho _{n+m}\left( \xi
,\xi ^{\prime }\right) -\rho _{n}\left( \xi \right) \rho _{m}\left( \xi
^{\prime }\right) .
\end{equation}%
Thus, we see that in order to calculate all possible field correlation functions of the form
$$
\left\langle 0\left\vert \varphi _{1},...\varphi _{n}\right\vert
0\right\rangle =\frac{\delta }{\delta J_{1}}...\frac{\delta }{\delta J_{n}}%
e^{W\left( J\right) }|_{J=0}
$$
it requires knowledge of the vacuum n-point distributions of wormholes $\rho_{n}\left( \xi_{1},...,\xi_{n}\right) $ defined by the expression (\ref{rho}).

\section{Euclidean action and background metric }

In quantum gravity , the Euclidean gravitational action is defined as \cite{H79}
\begin{equation}
S_{E}=-\frac{1}{16\pi }\int R\sqrt{g}d^{4}x-
\frac{1}{8\pi }\int K\sqrt{^3g}d^{3}x+
C  \label{Act1}
\end{equation}%
Here  $K$ is the trace of the second quadratic form on the boundary of the domain by which the action is calculated, $^3g_{ij}$ is the metric induced on the boundary, and we use Planck units in which $G=c=\hslash=1$. The constant $C$ is chosen so that in a flat Euclidean space the action is zero.

For each fixed topology of the space, the metric naturally decomposes into a background metric and perturbations. Of course, the background metric has the least Euclidean action, i.e., it obeys Einstein's Euclidean equations. As suggested in \cite{H87}, it is convenient to use a conformally flat metric for background geometry
$$
ds^{2}=h^{2}g_{ij}^{\ast }dx^{i}dx^{j}
$$
(in Cartesian coordinates $g_{ij}^{\ast }=\delta _{ij}$). Then the nontrivial topological structure will be defined in terms of the usual flat Euclidean space $x\in E$ by introducing additional boundary surfaces and corresponding boundary conditions.
As will be shown below, the real topological structure of the space is encoded in the background metric and is actually determined by a single scale function $h$. Deviations from the background metric or perturbations are assumed to be small and will be considered along with all other fields of matter.

For a conformally flat metric, the action reduces to the form
\begin{equation}
S_{E}(h)=-\frac{3}{8\pi }\int h\left(
-\Delta +\frac{1}{6}R^{\ast }\right) h\sqrt{g^{\ast }}d^{4}x
-\frac{3}{8\pi }\int h\nabla hd^{3}x  \label{Act}
\end{equation}%
where $\Delta =\square _{E}$ is the Laplacian with the metric $g_{ij}^{\ast} $. Thus, since in a flat Euclidean space the curvature is zero $R^{\ast}=0$, the Einstein equations in vacuum for the background metric are reduced to the Laplace equation
\begin{equation}\label{Lap}
-\Delta h=0,
\end{equation}%
which means that the conformal factor $h$ is a harmonic function. Since the real background topology is determined by the presence of additional boundary surfaces in $E$, this equation must be supplemented by the corresponding boundary conditions on such surfaces. We point out that the number of harmonic functions depends on the topology of the space (i.e., on the number of boundary surfaces and conditions on them), which means that the function $h$ contains explicitly all information about the topological structure of the space.

\section{Foam model as a set of Euclidean wormholes}

The most general nontrivial topology corresponding to a set of virtual wormholes is obtained as follows. Consider an ordinary Euclidean flat space $E$ and two copies of $\partial M_{\pm}$ of an arbitrary three-dimensional closed surface $\partial M$ in $E$. The part of the space $E$ that falls inside such surfaces $x\in M_{\pm}$ is removed, while the points on the corresponding two copies of the surfaces $\partial M_{\pm}$ are glued together. Thus, we get a Euclidean wormhole, the section of the neck of which is a 3-dimensional surface $\partial M$. Similarly, if we consider a set of copies of $\partial M_{\pm}^{A}$, where the index $A$ enumerates various closed 3-manifolds $\partial M^{A}$, and repeat the same procedure, we get a set of wormholes with neck sections $\partial M^{A}$. This procedure for obtaining an arbitrary topology of the space corresponds to the so-called Hegor diagrams \cite{book18,book19}, and the resulting space $E^{\ast} = E/M_{\pm}^{A}$ obtained from Euclidean space by removing part of the regions and the corresponding gluing, corresponds to a background space containing an arbitrary number of virtual (Euclidean) wormholes. As stated above, the background space metric is determined using the harmonic function $h$. In this case, gluing points on surfaces $\partial M^{A}$ sets specific periodic conditions on $h$.

Finding harmonic functions in the space of a general topological structure (i.e., in an ordinary Euclidean space in the presence of an arbitrary number of boundary surfaces of arbitrary shape) is a rather difficult task. However, the situation can be somewhat simplified. The fact is that the throats of wormholes $\partial M_{\pm}$ have a certain orientation in space $E^{\ast}$. Averaging over possible orientations, we lose some of the information, but as a result, instead of a closed 3-dimensional manifold $\partial M$ of a general form (in the sense of a topological structure)  we can already consider the usual three-dimensional spheres $\partial M\rightarrow S^{3}$.
Then the space $E^{\ast}$ is obtained simply by removing pairs of balls with subsequent gluing along the corresponding spheres, and the description of arbitrary topologies by a set of spherical wormholes can be considered as some leading approximation.

Note that the space $E^{\ast}$ can be viewed as an ordinary Euclidean space with a boundary. In other words, $E^{\ast}$ is an ordinary Euclidean space $E$ filled with a set of different three-dimensional spheres that form a boundary. The absence of a boundary surface in the space $E^{\ast}$ implies gluing over the corresponding pairs of boundary spheres. Accordingly, gluing sets specific boundary conditions for all fields specified on $E^{\ast}$.

In the case of spherical boundaries, the harmonic function $h$ defining the conformally-Euclidean metric can be obtained by the image method \cite{LL8}. In particular, in the case of three dimensions, a similar method is used in \cite{KS08}. Then the metric has the form
$ds^{2}=h^{2}ds_{\ast}^{2}$, where the conformal multiplier is a function of the form
\begin{equation}
 h=1+\sum_{\widetilde{A}}\frac{a_{\widetilde{A}%
}^{2}}{\left( x-R_{\widetilde{A}}\right) ^{2}}=1+\sum \frac{a_{A}^{2}}{%
\left( x-R_{A}\right) ^{2}}+\sum^{\prime }\frac{a_{b}^{2}}{\left(
x-R_{b}\right) ^{2}}  \label{GM}
\end{equation}%
where $R_ {\widetilde{A}}$ and $a_{\widetilde{A}}$ correspond to the positions and radii of pairs of balls or spheres forming a boundary in $E$ (i.e., which are removed from the space of $E^{\ast}$).
Note that from the point of view of the usual Euclidean space $E$, the scale function (\ref{GM}) obeys the Poisson equation
\begin{equation}
\label{Lap2}
-\Delta h=4\pi ^2  \sum_{\widetilde{A}}a_{\widetilde{A}
}^{2}\delta \left( x-R_{\widetilde{A}}\right)  .
\end{equation}%
However, all the sources on the right hand side of this equation correspond to the regions cut out of the original Euclidean space when obtaining the space $E^{\ast}$. Thus, the function $h$ is indeed harmonic on $E^{\ast}$ and obeys (\ref{Lap}).

The sum in (\ref{GM}) is divided into two parts. The first corresponds to real boundaries, i.e. a real set of initial balls with positions and radii $R_{A}$ and $a_{A}$, respectively. The second sum corresponds to the corrections that are obtained in the image method and which must be added so that $h$ satisfies the boundary conditions imposed by gluing.
It is important that all images (corrections) lie inside the original balls $\left(x-R_{A}\right) ^{2}<a_{A}^{2}$. These images are obtained by repeating inversion and gluing, as described in \cite{KS08}. Next, we describe a method for constructing an accurate metric for a set of wormholes. First, we will start with the simplest spherically symmetric wormhole that connects two asymptotically Euclidean spaces. Next, consider a generalization to a wormhole having both exits into the same space. Then we will generalize to an arbitrary case, i.e., an arbitrary set of wormholes with exits into the same Euclidean space.

\section{A set of wormholes with exits into the same Euclidean space}

The simplest configuration of a Euclidean wormhole was first considered by Hawking\cite{H87} and is described by the metric
\begin{equation}
ds^{2}=h^{2}(x)\delta _{\alpha \beta }dx^{\alpha }dx^{\beta },\,\,h=1+\frac{%
a^{2}}{(x^{\alpha }-x_{0}^{\alpha })^{2}}.  \label{X_wmetr}
\end{equation}%
This metric is invariant with respect to the transformation $T$ which is
\begin{equation}
x^{\mu \prime }-x_{0}^{\mu \prime }=\frac{a^{2}}{(x^{\alpha }-x_{0}^{\alpha
})^{2}}\Lambda _{\nu }^{\mu }(x^{\nu }-x_{0}^{\nu })  \label{inv}
\end{equation}%
and which corresponds to the composition of the inversion with respect  to the sphere of radius $a$ and the center $x_{0}^{\alpha}$ and the subsequent rotation $\Lambda_{\nu}^{\mu}\in O(4)$. In the new primed  coordinates $x^{\mu \prime } $, the metric has the same appearance (\ref{X_wmetr}). If we define $\rho^{2}=(x^{\alpha } - x_{0}^{\alpha})^{2}$ as a 4-dimensional radial coordinate ($0\leq\rho<\infty$) and introduce a new coordinate as $r=\rho -a^{2}/\rho$, where $ - \infty <r<\infty$, the metric reduces to a Bronnikov-Ellis type wormhole
\begin{equation}
ds^{2}=dr^{2}+\left( r^{2}+4a^{2}\right) d\Omega ^{2},
\end{equation}%
%%%%%%%%%%%%%%%%%
where $d\Omega ^{2}$ is the interval on a 3-sphere of unit radius (4-dimensional angular part). This metric describes two asymptotically Euclidean spaces $r\geq 0$ and $r\leq0$ connected by a neck whose section is a 3-sphere of radius $2a$ (corresponding to the points $r=0$).

A metric corresponding to a wormhole connecting regions in the same Euclidean space is obtained by identifying (or gluing) two spaces through an element of a group of movements. Let $R_{\pm}$ be the centers of two identical spheres of radius $a$ in the regions $r>0$ and $r<0$, and $x_{\pm}$ be the corresponding coordinates in these regions. Then gluing is the rule
\begin{equation}
x_{+}^{\mu }=R_{+}^{\mu }+\Lambda _{\nu }^{\mu }(x_{-}^{\nu }-R_{-}^{\nu }).
\label{gl-r}
\end{equation}%
In terms of common coordinates, the scale harmonic function that defines the metric in the leading approximation is
\begin{equation}
h\simeq h_{0}=1+\frac{a^{2}}{(x^{\alpha }-R_{+}^{\alpha })^{2}}+\frac{a^{2}}{%
(x^{\alpha }-R_{-}^{\alpha })^{2}},
\end{equation}%
where the physical domain corresponds to $(x^{\alpha }-R_{\pm }^{\alpha })^{2}\geq a^{2}$.
From the point of view of the space $E^{\ast}$, the inner regions of these spheres have been removed.

However, this function does not obey the continuity relations established by the gluing rule (\ref{gl-r}).
This disadvantage can be easily eliminated using the invariance of the metric with respect to the transformation of the type $T_{\pm }$ (\ref{inv}), defined with respect to each of the spheres.
Indeed, the inner regions $(x ^{\alpha }-R_{\pm }^{\alpha}) ^{2}<a^{2}$ must represent exact copies of the outer region, which means that the exact metric must be invariant under inversion and rotation relative to the spheres given by the equations $(x^{\alpha }-R_{\pm }^{\alpha })^{2}=a^{2}$.
Therefore, the exact scale function is obtained by iterating compositions of inversion  (\ref{inv}) and gluing (\ref{gl-r}). It is a series of additional images that can be numbered in order (or number) of iterations $h=\sum_{0}^{\infty}h_{n}$. The first iteration gives
\begin{equation}
h_{1}=\frac{a_{1}^{2}}{(x^{\alpha }-R_{1,+}^{\alpha })^{2}}+\frac{a_{1}^{2}}{%
(x^{\alpha }-R_{1,-}^{\alpha })^{2}},
\end{equation}%
where the new radii are $a_{1}^{2}=\frac{a^{2}}{X^{2}}a^{2}$, $X^{\alpha}=R_{+}^{\alpha }-R_{-}^{\alpha }$ is the distance between the spheres and
\begin{equation}
R_{1,\pm }^{\alpha }=R_{\pm }^{\alpha }\mp \frac{a^{2}}{X^{2}}\left( \Lambda
^{\pm 1}\right) _{\beta }^{\alpha }X^{\beta }.
\end{equation}%

It is important that the positions $R_{1,\pm }^{\alpha }$ lie inside the corresponding balls $(x^{\alpha }-R_{\pm }^{\alpha })^{2}<a^{2}$ and define new balls of radius $a_{1}^{2}$. Thus, we get an iterative procedure $(R_{n+1},,a_{n+1})=T\left( R_{n},a_{n}\right) $, etc. which generates subsequent images and corrections $h_{n}$. Note that in the case of a single wormhole, all images form a system of nested balls.
Generalization to the case of an arbitrary number of wormholes is
straightforward. Now we get a set of primary balls with parameters $a_{A}$, $R_{A,\pm},\Lambda_{A}$, which define the corresponding iterative mappings $T_{A}$ (relative to each specific throat) and generate a set of images of balls of the n-th order, applying maps of the form $T_{A}$, $T_{A}T_{B}$, ..., $\prod_{k=1}^n T_{A_k}$, etc. Thus, we obtain the final  metric of the form (\ref{GM}), which is invariant with respect to all mappings of $T_ {A}$ and their compositions.

Indeed, to see this, let's choose spherical coordinates around the center of a particular ball
\begin{equation}
ds^{2}=\left( 1+\frac{a^{2}}{r^{2}}+\sum \frac{a_{j}^{2}}{(x^{\alpha
}-x_{0j}^{\alpha })^{2}}\right) ^{2}\left( dr^{2}+r^{2}d\Omega ^{2}\right)
\end{equation}%
The inversion defines the coordinate transformation as follows
\[
r^{\prime }=\frac{a^{2}}{r},
%r=\frac{a^{2}}{r^{\prime 2}}r^{\prime },\ rr^{\prime }=a^{2},
\ dr=-\frac{a^{2}}{r^{\prime 2}}dr^{\prime }
\]%
what gives
\begin{equation}
ds^{2}=\left( 1+\frac{a^{2}}{r^{\prime 2}}+\sum \frac{a_{j}^{\prime 2}}{%
(x^{\alpha \prime }-x_{0j}^{\alpha \prime })^{2}}\right) ^{2}\left(
dr^{\prime 2}+r^{\prime 2}d\Omega ^{2}\right) ,
\end{equation}%
where
$$
\ a_{j}^{\prime 2}=\frac{a^{2}%
}{r_{0j}^{2}}a_{j}^{2}, \ r_{0j}^{\prime }=\frac{a^{2}}{r_{0j}^{2}}r_{0j}.
$$
To determine the mapping of $T$ relative to a given ball, we must add only the rotation $\Lambda _j$ of the resulting metric, which defines the new positions $r_{0j}^{\prime}$, but does not change the shape of the metric. Consequently, we see that the exact metric describing the gas of wormholes is invariant under such transformations. This means that the sum of $h=1+\sum_{\widetilde{A}}\frac{a_{\widetilde{A}}^{2}}{\left(x-R_{\widetilde{A}}\right) ^{2}}$ remains the same.

The image method implies that we first consider the actual number of wormholes with a set of parameters $a_{A}$, $R_{A}$, and then add all possible additional images that are obtained by inversions and rotations (according to the maps $T_{A}$) relative to all spheres with radii $a_{A}$ and the centers of $R_{A}$.
Repeating the iterations results in a series of images.
The first iteration creates additional images with new radius values $a_{b}^{2}=a_{A}^{2}a_{B}^{2}/X_{AB}^{2}$
and new positions $R_{b}$, in this case, $X_{AB}^{2}=\left(R_{A}-R_{B}\right) ^{2}$.
The next iteration gives balls of radii $a_{b}^{2}=a_{A}^{2}a_{B}^{2}a_{C}^{2}/(X_{AB}^{2}X_{AC}^{2})$, etc.

\section{Dependence of Euclidean action on wormhole parameters}

Let's now consider the value of the Euclidean action for an arbitrary number of wormholes. In a conformally flat space $R^{\ast}=0$, the function $h$ is harmonic, and the integral (\ref{Act}) reduces only to the surface term. It is important that the integration into (\ref{Act}) is performed over the areas of $\frac{a_{A}^{2}}{\left(x-R_{A}\right) ^{2}}\geq 1$. Therefore, each ball defines an additional boundary spherical surface $\left(x-R_{A}\right) ^{2}=a_{A}^{2}$.
We assume that all wormholes have both exits to the space $E^{\ast}$, and therefore such surfaces appear in pairs. Then the contribution of both pairs of surfaces compensates each other, and we have to take into account only a single boundary surface at infinity. Then we get
\begin{equation}
S_{E}=-\frac{3}{8\pi }\int h\nabla hd^{3}\Sigma =-\frac{3}{8\pi }2\pi
^{2}\lim_{R\rightarrow \infty }R^{3}h\nabla _{R}h
%=\frac{3}{8\pi }2\pi
%^{2}\sum \lim_{R\rightarrow \infty }R^{3}\frac{2a_{\widetilde{A}}^{2}}{R^{3}}%
=\frac{3\pi }{2}\sum a_{\widetilde{A}}^{2} .
\end{equation}%
Therefore, the action for an arbitrary set of wormholes is simply
\begin{equation}
S_{E}(h)=\frac{3}{2}\pi \sum_{\widetilde{A}}a_{\widetilde{A}}^{2}=\frac{3}{2}%
\pi \left[ \sum a_{A}^{2}+\sum_{AB}\frac{a_{A}^{2}a_{B}^{2}}{X_{AB}^{2}}+...%
\right] ,  \label{ActV}
\end{equation}%
where $\left(a_{A}\right)$ is a set of independent throats, while the set $\left( a_{\widetilde{A}}\right) $  contains all possible images. Here we have clearly written out only the first-order images. All subsequent terms are easily obtained using the iterative procedure for obtaining additional images described in the previous section.

\section{Vacuum distribution of wormholes}

Before considering the average values for n-point distributions, we will make the following remark.
The average vacuum values determined in (\ref{Mn}) are determined not only by the dependence of the action of $S_{E}(h)$ for the background metric on the parameters of wormholes, but also by the dependence of the function $Z_0(G)$ on these parameters.
For convenience, one can, in some approximation, first get rid of such dependence in the following way.
Using the representation $G=\overline{G}+\Delta G$ we write
\begin{equation}
Z_{0}\left( G\right) =Z_{0}\left( \overline{G}+\Delta G\right) =Z_{0}\left(
\overline{G}\right) \left( 1+\frac{Z_{0}\left( \overline{G}\right) ^{\prime }%
}{Z_{0}\left( \overline{G}\right) }\Delta G+...\right) .
\end{equation}%
Then, in the leading approximation, we can take $Z_{0}\left(G\right) \simeq Z_{0}\left(\overline{G}\right)$, while the rest of the corrections can be taken into account according to perturbation theory. Obviously, all such corrections are also expressed in terms of various averages of the type $\overline{(\Delta G)^n}$, which in turn are expressed in terms of the same distributions $\rho_{n}\left(\xi\right)$. In this case, $Z_{0}\left(\overline{G}\right) $ does not depend on the parameters of specific wormhole distributions (it depends only on the average vacuum distribution) and it can be taken into account in the normalization constant.

Thus, we define
\begin{equation}\label{r1}
\rho _1\left( \xi \right) =\left\langle 0\left\vert \sum_{A}\delta \left( \xi
-\xi _{A}\right) \right\vert 0\right\rangle =\frac{1}{Z}\int F\left( \xi
\right) e^{-S_{E}\left( h\right) }DF,
\end{equation}%
where $DF=\prod d\xi _{A}$ and $S_{E}(h)$ is given by (\ref{ActV}), and the normalization constant is given by the integral
$$
Z=\int e^{-S_{E}}DF.
$$
Similarly, we define a two-point distribution function as
\begin{equation}\label{r2}
\rho _2\left( \xi ,\xi ^{\prime }\right) =\left\langle 0\left\vert
\sum_{AB}\delta \left( \xi -\xi _{A}\right) \delta \left( \xi ^{\prime }-\xi
_{B}\right) \right\vert 0\right\rangle =\frac{1}{Z}\int F\left( \xi \right)
F\left( \xi ^{\prime }\right) e^{-S_{E}}DF.
\end{equation}%
Using the identity
\[
\sum_{A,B}\delta \left( \xi -\xi _{A}\right) \delta \left( \xi
^{\prime }-\xi _{B}\right) =\delta \left( \xi ^{\prime }-\xi \right)
\sum_{A}\delta \left( \xi -\xi _{A}\right) +\sum_{A\neq B}\delta
\left( \xi -\xi _{A}\right) \delta \left( \xi ^{\prime }-\xi _{B}\right)
\]
we get an expression for a two-point function in the form
\begin{equation}
\rho _2 \left( \xi ,\xi ^{\prime }\right) =\delta \left( \xi ^{\prime }-\xi
\right) \rho _1 \left( \xi \right) +\left\langle \sum_{A\neq B}\sum_{B}\delta
\left( \xi -\xi _{A}\right) \delta \left( \xi ^{\prime }-\xi _{B}\right)
\right\rangle .
\end{equation}%
Then representing the expression (\ref{ActV}) as
\begin{equation}\label{SU}
S_{E}(h)= \sum _A S_{0}(\xi _A)+\sum_{A\neq B}U( \xi _A,\xi _B)+...
\end{equation}%
and considering the second sum and $U(\xi _A,\xi _B)$, as a small perturbation we find
\begin{equation}
\rho _1 \left( \xi \right) =\rho _{0}\left( \xi \right) \left( 1-\Phi \left(
\xi \right) +...\right) ,
\end{equation}%
\begin{equation}
\Delta \rho _{1,1}\left( \xi ,\xi ^{\prime }\right)  =\delta \left( \xi ^{\prime }-\xi \right) \rho _1\left(
\xi \right) -\rho _1\left( \xi \right) \rho _1\left( \xi ^{\prime }\right) \left[
2U\left( \xi ,\xi ^{\prime }\right)  + \dots
\right]
\end{equation}%
where $\Delta\rho _{1,1}\left(\xi ,\xi ^{\prime }\right)$ is defined in (\ref{DR}), and $\rho_{0}\left(\xi \right)$ corresponds to the approximation of the "ideal gas of wormholes".  This approximation corresponds to the case when only the first sum is retained in (\ref{ActV}), and we neglect the terms responsible for mutual influence or interaction between different wormholes, i.e., $U(\xi _A,\xi _B)\simeq 0$. The explicit calculation gives the expression
\begin{equation}
\rho _{0}\left( \xi \right) =N\frac{1}{z_{0}}e^{-2S_{0}\left( \xi \right) }.
\end{equation}%
Here $N$ is the number of wormholes, $S_{0}\left(\xi_{A}\right) $ is defined above in (\ref{SU}) (explicitly this function is given as $S_{0}\left(\xi_{A}\right) =$ $\frac{3}{2}\pi a_{A}^{2}\left[ 1+\frac{a_{A}^{2}}{X_{A}^{2}}+...\right] $), and $z_{0}$ is the normalization constant ($\int\frac{1}{z_{0}}e^{-2S_{0}\left( \xi
\right) }d\xi =1$).
The function $\Phi\left(\xi\right)$ is expressed in terms of the potential $U$ introduced in (\ref{SU}) as follows
\begin{equation}
\Phi \left( \xi \right) =\int  2U\left(
\xi ,\xi _{C}\right)  \rho _{0}\left( \xi _{C}\right) d\xi _{A}.
\end{equation}%
The explicit expression for the potential is $U\left(\xi ,\xi^{\prime }\right) $ is given by the formula
$$
U\left( \xi ,\xi ^{\prime }\right) =3\pi a^{2}a^{\prime 2}\left[ \frac{1}{%
\left( R_{+}-R_{+}^{\prime }\right) ^{2}}+\frac{1}{\left(
R_{+}-R_{-}^{\prime }\right) ^{2}}+\frac{1}{\left( R_{-}-R_{+}^{\prime
}\right) ^{2}}+\frac{1}{\left( R_{-}-R_{-}^{\prime }\right) ^{2}}\right] .
$$

We point out that all higher $n$-point distribution functions of interest are determined in a similar way.

\section{Concluding remarks }

Thus, we see that all $n$-point wormhole distribution functions in a vacuum allow analytical calculation and are entirely determined by the action (\ref{ActV}), (\ref{SU}). The approximation of a rarefied or ideal gas of wormholes corresponds to the case when only the first sum is preserved in (\ref{ActV}), and we neglect the terms responsible for the mutual influence or interaction between different wormholes.
In this approximation, the effective action for scalar particles was calculated in \cite{KSPV}. Taking into account the interaction between wormholes leads, apparently, to additional and significant changes.

First, we note that the influence of wormholes on each other leads to effective repulsion between the necks.  In other words, the presence of one wormhole in a given volume element of the configuration space leads to a decrease in the probability of detecting another wormhole in a given volume element. This means that the correction to an effective action will have the opposite sign.
As suggested in \cite{KSPV}, such a correction is described by a term of the form
$$
V_3(\phi )=-\frac{\pi^4}{2}\int a^2(\phi (R_+)-\phi (R_-))^2 a^{\prime 2}(\phi (R_+ ^{\prime })-\phi (R_- ^{\prime }))^2\omega (\xi, \xi ^{\prime })d\xi d\xi ^{\prime },
$$
where
$
\omega\left( \xi ,\xi ^{\prime }\right)  = -2U\left( \xi ,\xi ^{\prime }\right)\rho _1\left( \xi \right) \rho _1\left( \xi ^{\prime }\right) .
$

Secondly, we note that the naive estimate based on the homogeneous distribution of virtual wormholes in space apparently stops working. The homogeneous distribution of wormholes in space leads to too large an amount of the Euclidean action and in this sense becomes unstable. The situation here is quite similar to the instability of the homogeneous distribution of baryons in the modified MOND-type theory of gravity considered in \cite{K06}. Indeed, if you allocate a certain amount of space and fix the number of wormholes in it, then with a homogeneous distribution of wormholes, adding even one additional wormhole to a given volume leads to a significant increase in the magnitude of the action (\ref{ActV}). This is due to all the auxiliary images that appear, both on the necks of existing wormholes, and on the neck of the additional wormhole itself. All such images also make a positive contribution to the action (\ref{ActV}). This means that such distributions will be strongly suppressed and will not contribute to the average values of the form (\ref{r1}) and (\ref{r2}). At the same time, as shown in \cite{K06}, with a fractal distribution of initial sources in (\ref{Lap2}) (i.e. wormholes), adding one wormhole does not lead to a significant increase in action, as it should be for a stable distribution.
In other words, for a fixed number of wormholes, the fractal distribution of wormholes, rather than a homogeneous one, has the least Euclidean action.

\end{document}